**Synchrotron X-ray study of polycrystalline wurtzite $Zn_{1-x}Mg_xO$ ($0 \leq x \leq 0.15$): Evolution of crystal structure and polarization**

Young-Il Kim, Katharine Page, and Ram Seshadri

*Materials Department and Materials Research Laboratory, University of California, Santa Barbara, California, 93106, USA*

ykim@mrl.ucsb.edu, kpage@mrl.ucsb.edu, seshadri@mrl.ucsb.edu

The effect of Mg-substitution on the crystal structure of wurtzite ZnO is presented based on synchrotron X-ray diffraction studies of polycrystalline $Zn_{1-x}Mg_xO$ ($0 \leq x \leq 0.15$). Increase in Mg concentration results in pronounced *c*-axis compression of the hexagonal lattice, and in diminution of the off-center cation displacement within each tetrahedral $ZnO_4$ unit. Going from ZnO to $Zn_{0.85}Mg_{0.15}O$, significant changes in the ionic polarization are observed (−5.6 to −4.8 $\mu C/cm^2$), despite only subtle increments in the cell volume (~0.03 %) and the *ab*-area dimension (~0.1 %). The optical properties of the samples have also been characterized and the band gap changes from 3.24 eV (ZnO) to 3.35 eV ($Zn_{0.85}Mg_{0.15}O$).

From both fundamental and application viewpoints, zinc oxide (ZnO) is a unique material with diverse functions deriving from its semiconducting, piezoelectric, pyroelectric, photoluminescent, and photocatalytic characteristics.[1,2] Especially owing to the recent progresses in large area epitaxy and single crystal growth, ZnO is emerging as a promising material in the next-generation optoelectronic devices.[1–3] At room temperature ZnO has a direct band gap of 3.3 eV with a high exciton binding energy ~60 meV, is resistant to electron irradiation, can be processed by wet-chemical etching, and native substrates are available for well-controlled homoepitaxy.[1–3] In these respects, ZnO is considered as comparable or even superior to GaN or SiC as the active component in light emitting devices.

In analogy with the AlN-GaN-InN systems, alloys of $Zn_{1-x}Mg_xO$ and $Zn_{1-x}Cd_xO$ have been receiving attention[4,5] from the viewpoint of band gap engineering to realize the emission/detection of three primary colors of the visible spectrum. Of equal interest are heterojunctions of ZnO with $Zn_{1-x}Mg_xO$, which provide suitable architectures for laser diodes and polarization-doped field-effect transistors (PolFET).[6] Being known for its notable spontaneous polarization among the largest in II-VI or III-V semiconductors[7,8] wurtzite ZnO is appropriate for PolFETs. Fabrication of GaN based PolFETs have already been demonstrated using $Ga_{1-x}Al_xN$ sublayers, with improved performance over the traditional impurity-doped metal-semiconductor field-effect transistors (MESFET).[6] In modeling ZnO based PolFETs, $Zn_{1-x}Mg_xO$ is regarded as the most promising material for interfacing with ZnO, since it may allow near perfect lattice match as well as facile control of polarization gradients.

In recent years there has been significant research effort on $Zn_{1-x}Mg_xO$ using various sample types including films,[4,5,9,10] nanostructures,[11,12] polycrystalline powders,[13] and single crystals.[14] Although these studies have established obvious relationships between the Mg content and the optical properties of $Zn_{1-x}Mg_xO$, details of structural evolution have not so far been elucidated. Reported lattice constants of $Zn_{1-x}Mg_xO$ are mostly from film phases which are subject to lattice strains imposed by substrate. The only single crystal work on $Zn_{1-x}Mg_xO$[14] reported variations of lattice constants inconsistent with the trends observed from the film samples.[4,5,9,10] On the other hand, oxygen positions in $Zn_{1-x}Mg_xO$, which are of crucial importance to the polarization, have not been discussed by any authors. It is therefore an interesting task to analyze the detailed crystal structure of $Zn_{1-x}Mg_xO$ using fully equilibrated bulk samples. In this work we present the results of a high-flux/high-energy synchrotron X-ray diffraction study on polycrystalline $Zn_{1-x}Mg_xO$ solid solutions. It is shown that as a consequence of Mg-substitution the static polarization in the crystal, calculated using point charge models, can be gradually varied, in parallel with the internal tetrahedral distortion.

Polycrystalline samples of $Zn_{1-x}Mg_xO$ ($x$ = 0, 0.05, 0.10, 0.15, and 0.20) were prepared by an oxalate-based coprecipitation method.[15] Aqueous solutions of Zn and Mg acetates were mixed in $H_2C_2O_4$ solution in the ratio of $[Zn^{+2}]:[Mg^{+2}]:[C_2O_4^{-2}]$ = (1−$x$):$x$:1.05, to coprecipitate zinc magnesium oxalates. The precipitates were washed with deionized water and dried at 60°C for 4 h to produce white powders of $Zn_{1-x}Mg_x(C_2O_4)\cdot 2H_2O$. For all the $Zn_{1-x}Mg_x(C_2O_4)\cdot 2H_2O$ samples with 0 ≤ $x$ ≤ 0.15, thermogravimetry in air (Cahn ThermMax 400 TGA, heating up to 1000°C at 5°C/min) clearly showed the stepwise processes of dehydration at ~150°C and oxalate-to-oxide conversion at ~390°C, with the weight changes in excellent agreement with expected values. The oxalate dihydrates were transformed to $Zn_{1-x}Mg_xO$ by heating in air at 550°C for 24 h.

X-ray powder diffraction (XRPD) measurements (Phillips X'PERT MPD, Cu K$\alpha_{1,2}$ radiation with 45 kV/ 40 mA) confirmed phase formation of $Zn_{1-x}Mg_x(C_2O_4)\cdot 2H_2O$,[16] and wurtzite-type $Zn_{1-x}Mg_xO$. However on the sample with $x = 0.20$ ($Zn_{0.80}Mg_{0.20}O$), a weak impurity peak was observed at $2\theta \approx 43°$ presumably due to the 200 diffraction of cubic MgO. In an earlier report[17] the thermodynamic solubility of MgO in ZnO was indicated as 2 wt% ($\approx 4$ mol%), but we could not detect any segregation of MgO phase from the samples with $x \leq 0.15$. Fourier-transform infrared (FT-IR) spectra of $Zn_{1-x}Mg_xO$ powders were recorded in KBr using a Nicolet Magna 850 FT-IR spectrophotometer in the transmission mode. A previous report pointed out that organic precursors for ZnO synthesis may leave carbonate species strongly bound within the lattice,[18] but the samples in this study did not exhibit any spectral feature at ~1300 and ~1500 cm$^{-1}$ (ref 19) demonstrating the complete decomposition of oxalate. The wurtzite lattice vibrations were observed as broad IR bands at 400–600 cm$^{-1}$. Upon Mg-substitution, these stretching modes shifted to higher wavenumber, attributed to the smaller reduced mass of Mg–O. Diffuse-reflectance absorption spectra were measured on a Shimadzu UV-3600 spectrophotometer equipped with an ISR-3100 integrating sphere for $Zn_{1-x}Mg_xO$ in the wavelength range of 220–800 nm. The optical band gap was determined by extrapolating the absorption edge to zero-absorption. Depending on the Mg content $x$, the band gap energy gradually increased from 3.24 ($x = 0$) to 3.26 ($x = 0.05$), 3.30 ($x = 0.10$), and 3.35 eV ($x = 0.15$). These values can be compared with previous reports on thin films; 3.36 ($x = 0$), 3.63 ($x = 0.14$), and 3.87 eV ($x = 0.33$).[4] In any cases, samples of ZnO with vastly differing optical properties are known to have indistinguishable structures.[20] Crystal structures of $Zn_{1-x}Mg_xO$ were analyzed by Rietveld refinements of the XRPD patterns collected using synchrotron radiation ($\lambda \sim 0.137$ Å) at beam line 11-ID-B of Advanced Photon Source, Argonne National Laboratory. The synchrotron wavelength was carefully calibrated to 0.13648 Å by using the cell constants of ZnO determined from Cu K$\alpha$ radiation. Sample powders were loaded in Kapton tubes and the data were measured in transmission mode using an amorphous silicon detector from General Electric Healthcare. The program FIT2D[21] was employed to process the images to the corresponding one-dimensional XRPD pattern. For Rietveld refinements, the GSAS-EXPGUI software[22] suite was employed.

Detailed structure analyses were performed for $Zn_{1-x}Mg_xO$ ($x$ = 0, 0.05, 0.10, and 0.15), by Rietveld refinement over 2θ ranges of 1.5–22.2° ($d$ > 0.355 Å). Figure 1 shows data on ZnO and $Zn_{0.85}Mg_{0.15}O$, together with the Rietveld refinement profiles. The substitution of Mg did not cause marked changes in the diffraction patterns as expected from the similar four-coordination ionic radii[23] of $Zn^{+2}$ (0.60 Å) and $Mg^{+2}$ (0.57 Å). The structure model was taken from wurtzite ZnO in space group of $P6_3mc$ and both Zn/Mg and O at 2$b$ Wyckoff positions, (1/3, 2/3, 0) and (1/3, 2/3, $u$), respectively.[24] For all atom types isotropic temperature factors were refined, and Zn/Mg were statistically distributed over the common crystallographic site. Due to non-negligible correlation effects, the occupancies of Zn and Mg in $Zn_{1-x}Mg_xO$ ($x$ > 0) could not be simultaneously refined with the scale factor, but once scale factors were fixed, the occupancies converged to $Zn_{0.950(1)}Mg_{0.050(1)}O$, $Zn_{0.899(1)}Mg_{0.101(1)}O$, and $Zn_{0.850(1)}Mg_{0.150(1)}O$. In all four samples, the final Rietveld $R_{wp}$ values were obtained as ≤ 4 %.

From the refined structural parameters (Table I and Figure 2), it is observed that Mg substitution results in an elongation of the $a$-axis and a contraction of the $c$-axis. The overall consequence is a more pronounced wurtzite distortion, where the $ZnO_4$ tetrahedra are uniformly compressed along the $c$-axis. While the parent ZnO is already substantially distorted, as indicated by the deviation of $c/a$ ratio (1.6021) from that of an ideal geometry 1.633, the hexagonal lattice is further deformed upon Mg substitution. From existing wurtzite structures, it is well known that when the bonding character becomes more ionic, the $c/a$ ratio moves further from the ideal value.[25,26] Some previous studies on film phases have reported similar dependences of $a$- and $c$-parameters,[4,5,9,10] but as noted previously, the films may experience strain from the substrate lattice.

Another important distortion in the wurtzite structure arises from the $c$-axis cation displacement, which is measured by the deviation of the anion positional parameter $u$ from an ideal value of 0.375. The four nearest cation-anion pairs are equidistant when $u = a^2/(3c^2) + 0.25$, whereas the dipole moments within the each $ZnO_4$ tetrahedron is zero if $u$ = 0.375, regardless of the $c/a$ ratio. As plotted in Figure 2, the $u$ parameter in $Zn_{1-x}Mg_xO$ solid solution approaches 0.375 as $x$ increases, which in turn results in more regular inter-atomic distances and angles. Since there are only a few wurtzite structures known with atomic parameters, it is difficult to find an empirical dependence of $u$ on the $c/a$ ratio or the bonding character. However, it can be qualitatively stated that $u$ tends to vary in such a way that the four tetrahedral distances remain as constant as possible. Figure 3 indicates that the existing wurtzite compounds roughly maintain the relationship of $u = a^2/(3c^2) + 0.25$. ZnO is found with the highest $u$ among the examples, and also with the $c/a$ being extremely deviated from the ideal value. The tetrahedral distortion is partially relieved by alloying with MgO. In ZnO the cation is displaced from the center of zero-polarization by 0.041 Å and such displacement is lessened to 0.036 Å in $Zn_{0.85}Mg_{0.15}O$.

We have calculated the static polarization $P_s$ along 001, using a point charge model and taking as reference (0 μC/cm$^2$) the structure with zero dipoles within tetrahedra. It monotonically decreases from −5.6 μC/cm$^2$ for ZnO to −4.8 μC/cm$^2$ for $Zn_{0.85}Mg_{0.15}O$. For comparison density functional *ab initio* studies have suggested the $P_s$ of ZnO as −5 μC/cm$^2$ (refs 7,28) and −5.7 μC/cm$^2$ (refs 8,29) similar to the above ionic charge model. However, the corresponding experimental report is rarely found and the only experimental value −7±2 μC/cm$^2$ was deduced from second harmonic generation measurements.[30] Gopal and Spaldin have studied polarization properties of hypothetical wurtzite MgO structures. In the case the geometry is optimized, the resulting MgO wurtzite is predicted to have a much larger $P_s$ than the relaxed ZnO structure (−17 vs. −5 μC/cm$^2$) which does not agree with the trend observed here. In another case where Mg simply replaces Zn with the other wurtzite parameters unchanged, they predict that the $P_s$ of crystal will be lowered by ~8 %.[28] The latter result reflects the distinct effective charges of $Zn^{+2}$ and $Mg^{+2}$, and implies that if the smaller electronic polarizability of $Mg^{+2}$ is taken into account, the polarization gradient between ZnO and $Zn_{1-x}Mg_xO$ will be greater than the approximation made here from simple point charge considerations.

The $Zn_{1-x}Mg_xO$ samples in this study represent dilute MgO solutions in ZnO, where the structure parameters are nearly linear functions of $x$. Since the $P_s$ in wurtzite structure is very sensitive to $u$ parameter, the polar behavior of $Zn_{1-x}Mg_xO$ can be rationally controlled through composition. At present, epitaxial superlattices of $ZnO/Zn_{1-x}Mg_xO$ are fairly well established[31] and the fabrication of ZnO based PolFETs will no doubt be realized in the near future.

We gratefully acknowledge discussions with Nicola Spaldin, and support from the National Science Foundation through the MRSEC program, DMR05-20415. KP has been supported by the NSF through IGERT and Graduate Student Fellowships. Work at Argonne National Labs and the Advanced Photon Source was supported by DOE Office of Basic Energy Sciences, under contract W-31-109-Eng.-38. We thank Peter Chupas and Karena Chapman for their help with synchrotron data collection.[1] U. Ozgur, Ya. I. Alivov, C. Liu, A. Teke, M. A. Reshchikov, S. Dogan, V. Avrutin, S.-J. Cho, and H. Morkoc, J. Appl. Phys. **98**, 041301 (2005).

[2] S. J. Pearton, D. P. Norton, K. Ip, Y. W. Heo, and T. Steiner, J. Vac. Sci. Tech. B **22**, 932 (2004); S. J. Pearton, D. P. Norton, K. Ip, Y. W. Heo, and T. Steiner, Prog. Mater. Sci. **50**, 293 (2005).

[3] D. C. Look, J. Electron. Mater. **35**, 1299 (2006); Mater. Res. Eng. B **80**, 383 (2001).

[4] A. Ohtomo, M. Kawasaki, T. Koida, K. Masubuchi, H. Koinuma, Y. Sakurai, Y. Yoshida, T. Yasuda, and Y. Segawa, Appl. Phys. Lett. **72**, 2466 (1998).

[5] T. Makino, Y. Segawa, M. Kawasaki, A. Ohtomo, R. Shiroki, K. Tamura, T. Yasuda, and H. Koinuma, Appl. Phys. Lett. **78**, 1237 (2001).

[6] S. Rajan, H. Xing, S. DenBaars, U. K. Mishra, and D. Jena, Appl. Phys. Lett. **84**, 1591 (2004).

[7] A. Dal Corso, M. Posternak, R. Resta, and A. Baldereschi, Phys. Rev. B **50**, 10715 (1994).

[8] F. Bernardini, V. Fiorentini, and D. Vanderbilt, Phys. Rev. B **56**, 10024 (1997).

[9] X. Zhang, X. M. Li, T. L. Chen, C. Y. Zhang, and W. D. Yu, Appl. Phys. Lett. **87**, 092101 (2005).

[10] W. I. Park, G.-C. Yi, and H. M. Jang, Appl. Phys. Lett. **79**, 2022 (2001).

[11] W. Q. Peng, S. C. Qu, G. W. Cong, and Z. G. Wang, Appl. Phys. Lett. **88**, 101902 (2006).

[12] C.-H. Ku, H.-H. Chiang, and J.-J. Wu, Chem. Phys. Lett. **404**, 132 (2005).

[13] M. S. Tomar, R. Melgarejo, P. S. Dobal, and R. S. Katiyar, J. Mater. Res. **16**, 903 (2001).

[14] B. Wang, M. J. Callahan, and L. O. Bouthillette, Cryst. Growth Design **6**, 1256 (2006).

[15] A. S. Risbud, N. A. Spaldin, Z. Q. Chen, S. Stemmer, and R. Seshadri, Phys. Rev. B **68**, 205202 (2003)

[16] L. Guo, Y. Ji, H. Xu, Z. Wu, and P. Simon, J. Mater. Chem. **13**, 754 (2003).

[17] J. F. Sarver, F. L. Katnack, and F. A. Hummel, J. Electrochem. Soc. **106**, 960 (1959).

[18] W. M. Hlaing Oo, M. D. McCluskey, A. D. Lalonde, and M. G. Norton, Appl. Phys. Lett. **86**, 073111 (2005).

[19] J. Saussey, J.-C. Lavalley, and C. Bovet, J. Chem. Soc., Faraday Trans. **78**, 1457 (1982).

[20] H. Sawada, R. Wang, and A. W. Sleight, J. Solid State Chem. **122**, 148 (1996).

[21] A. P. Hammersley, S. O. Svensson, M. Hanfland, A. N. Fitch, and D. Hausermann, High Pressure Res. **14**, 235 (1996).

[22] A. C. Larson and R. B. von Dreele, *General Structure Analysis System*; Los Alamos National Laboratory Report LAUR 86-748 (1994); B. H. Toby, J. Appl. Cryst. **34**, 210 (2001).

[23] R. D. Shannon, Acta Cryst. A **32**, 751 (1976).

[24] S. C. Abrahams and J. L. Bernstein, Acta Cryst. B **25**, 1233 (1969).
6                                                          (11/15/2006)

[25] H. Schulz and K. H. Thiemann, Solid State Commun. **23**, 815 (1977).

[26] G. A. Jefferey, G. S. Parry, and R. L. Mozzi, J. Chem. Phys. **25**, 1024 (1956); F. Keffer and A. M. Portis, J. Chem. Phys. **27**, 675 (1957); P. Lawaetz, Phys Rev. B **5**, 4039 (1972).

[27] ZnO: ref 24; AlN, GaN: ref 25; ZnS: E. H. Kisi and M. M. Elcombe, Acta Cryst. C **45**, 1867 (1989); AgI: G. Burley, J. Chem. Phys. **38**, 2807 (1963); CdSe: A. W. Stevenson and Z. Barnea, Acta Cryst. B **40**, 530 (1984); CdS: A. W. Stevenson, M. Milanko, and Z. Barnea, Acta Cryst. B **40**, 521 (1984); BeO: J. W. Downs, F. K. Ross, and G. V. Gibbs, Acta Cryst. B **41**, 425 (1985); SiC: H. Schulz and K. H. Thiemann, Solid State Commun. **32**, 783 (1979).

[28] P. Gopal and N. A. Spaldin, J. Electron. Mater. **35**, 538 (2006).

[29] Y. Noel, C. M. Zicovich-Wilson, B. Civalleri, Ph. D'Arco, and R. Dovesi, Phys. Rev. B **65**, 014111 (2001).

[30] J. Jerphagnon and H. W. Newkirk, Appl. Phys. Lett. **18**, 245 (1971); R. C. Miller, Appl. Phys. Lett. **5**, 17 (1964).

[31] Th. Gruber, C. Kirchner, R. Kling, F. Reuss and A. Waag, Appl. Phys. Lett. **84**, 5359 (2004); A. Ohtomo, M Kawasaki, I. Ohkubo, H. Koinuma, T. Yasuda, and Y. Segawa, Appl. Phys. Lett. **75**, 980 (1999).

TABLE I. Structural parameters for polycrystalline $Zn_{1-x}Mg_xO$ determined by the Rietveld refinement of synchrotron XRPD data in space group $P6_3mc$ with Zn/Mg at (1/3, 2/3, 0) and O at (1/3, 2/3, $u$).

|  | | ZnO (ref. 24) | \multicolumn{4}{c}{$x$ in $Zn_{1-x}Mg_xO$ (This work)} |
|---|---|---|---|---|---|---|
|  | | | 0 | 0.05 | 0.10 | 0.15 |
| $a$ (Å) | | 3.249858(6) | 3.25030(9) | 3.25088(8) | 3.25144(8) | 3.25208(8) |
| $c$ (Å) | | 5.206619(2) | 5.2072(2) | 5.2067(2) | 5.2052(2) | 5.2033(2) |
| Vol (Å$^3$) | | 47.622830(9) | 47.642(3) | 47.654(3) | 47.656(3) | 47.658(3) |
| $u$ | | 0.3825(14) | 0.3829(4) | 0.3826(4) | 0.3823(4) | 0.3819(4) |
| $U_{iso}$ (Å$^2$) | Zn/Mg | 0.0080(3) | 0.00577(6) | 0.00589(6) | 0.00582(6) | 0.00573(6) |
| | O | 0.0086(9) | 0.0059(5) | 0.0063(4) | 0.0075(4) | 0.0087(4) |

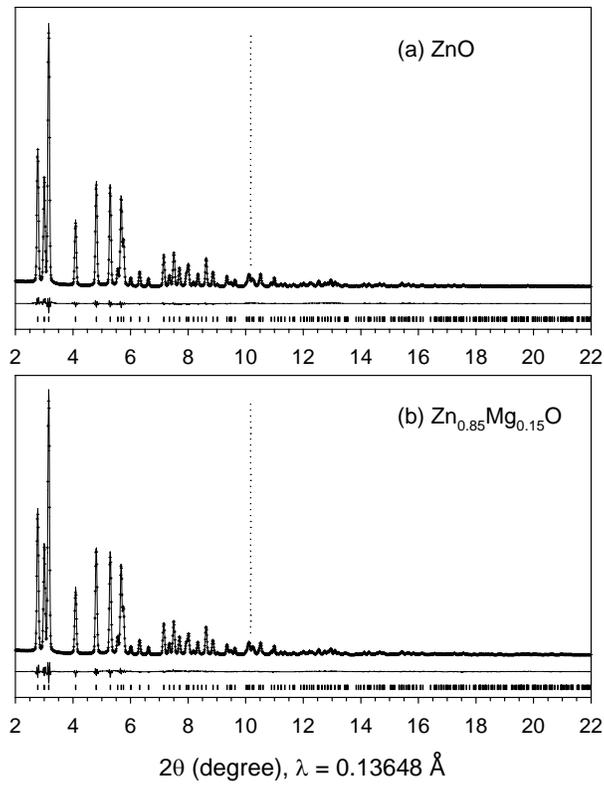

FIG. 1. Rietveld refinements of synchrotron ($\lambda = 0.13648$ Å) XRPD patterns for (a) ZnO and (b) $Zn_{0.85}Mg_{0.15}O$. Observed (crosses) and calculated (solid lines) data are overlapped, with the difference pattern and the expected peak positions shown at the bottom. Vertical dotted lines at $2\theta = 10.16°$ mark the $d$-spacing of 0.770 Å, which corresponds to $2\theta = 180°$ for Cu K$\alpha_1$ radiation.

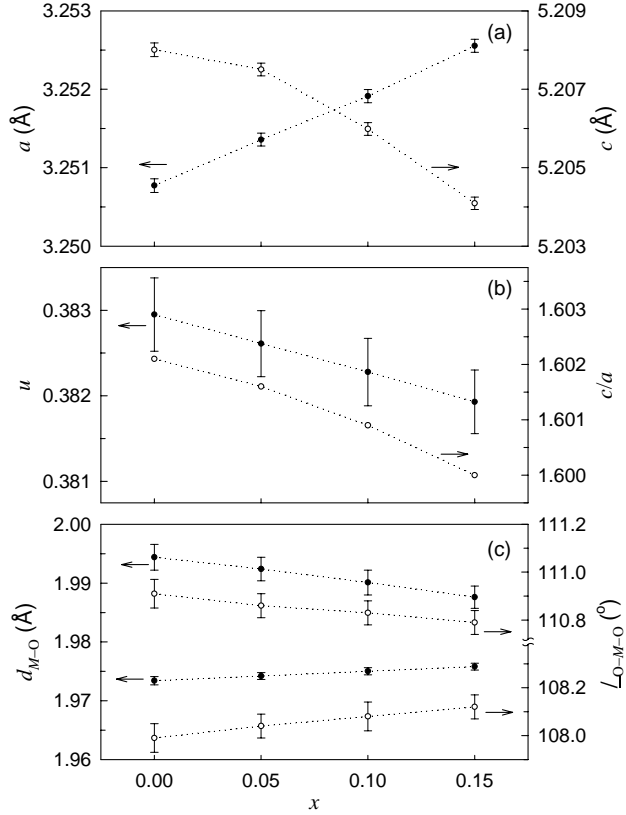

FIG. 2. Structural parameters depending on $x$ in $Zn_{1-x}Mg_xO$; (a) lattice constants, (b) $u$ and $c/a$ ratio, and (c) bond lengths and angles, where $M$ represents Zn and Mg which were not distinguished crystallographically. In each panel, filled and open circles are associated with the left and right axes, respectively.

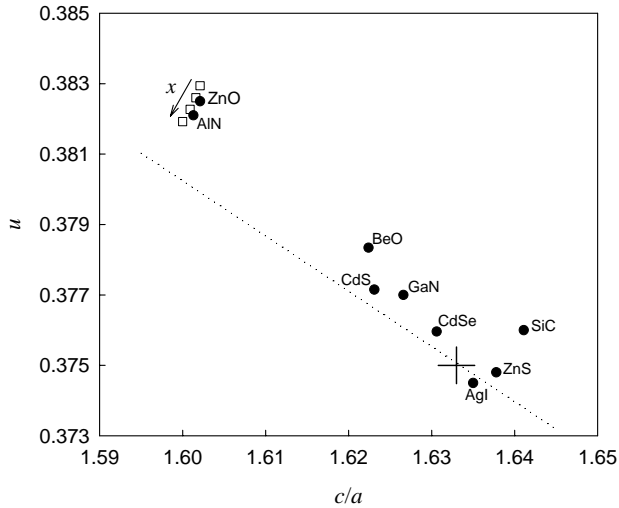

FIG. 3. Comparison between $u$ and $c/a$ parameters of experimental wurtzite structures. $Zn_{1-x}Mg_xO$ phases are marked by open squares ($x$ = 0, 0.05, 0.10, and 0.15 along the direction of the arrow), and the literature data,[27] filled circles. The dotted line corresponds to structures with equal bond lengths, and the crosshair represents the point of $c/a = 1.633$ and $u = 0.375$.